# Gravitational influence of Saturn's rings on its moons: a case for free granular flow

Troy Shinbrot

*Dept. of Physics & Astronomy, Rutgers University, Piscataway, NJ USA*

*Exploratory missions have found that regolith on interplanetary bodies can be loosely packed and freely flowing – a state that strongly affects mission plans and that may also influence the large-scale shapes of these bodies. We investigate here whether notable circumferential ridges seen on Saturn's moons may be a byproduct of free flow of loosely packed regolith. Such ridges and other features likely record the history of the moons, and we find that if surface grains are freely flowing, then the combined gravity of Saturn itself and its tenuous ring generate similar circumferential features. Moreover, analysis of these features reveals the possibility of previously unreported morphologies, for example a stationary torus around a non-rotating satellite. Some of these features persist even for a very low density and distant disk, which raises the prospect that nonlinear analysis of interactions from disks to moons and back again may lead to new insights.*

## Introduction:

International exploratory missions have visited interplanetary bodies and have found their regolith to be loosely packed and freely flowing [1,2]. These findings concur with both modeling [3] and terrestrial experiments [4,5] intended to investigate how the processes of fine scale regolith accretion produce larger scale morphogenesis. Nowhere is this morphogenesis more striking than on Saturn's moons, which exhibit remarkable equatorial features, some of which are shown in Fig. 1. Pan and Atlas (panels a & b) have such prominent equatorial ridges that the moons resemble ravioli, with aspect ratios around ½. Iapetus (panel c) is much larger and is nearly spherical, but has a sharp, 20 km high, ridge over most of its circumference. By contrast, the asteroid Vesta (panel d) is scoured by multiple, 2 km deep, troughs extending along much of its equator.

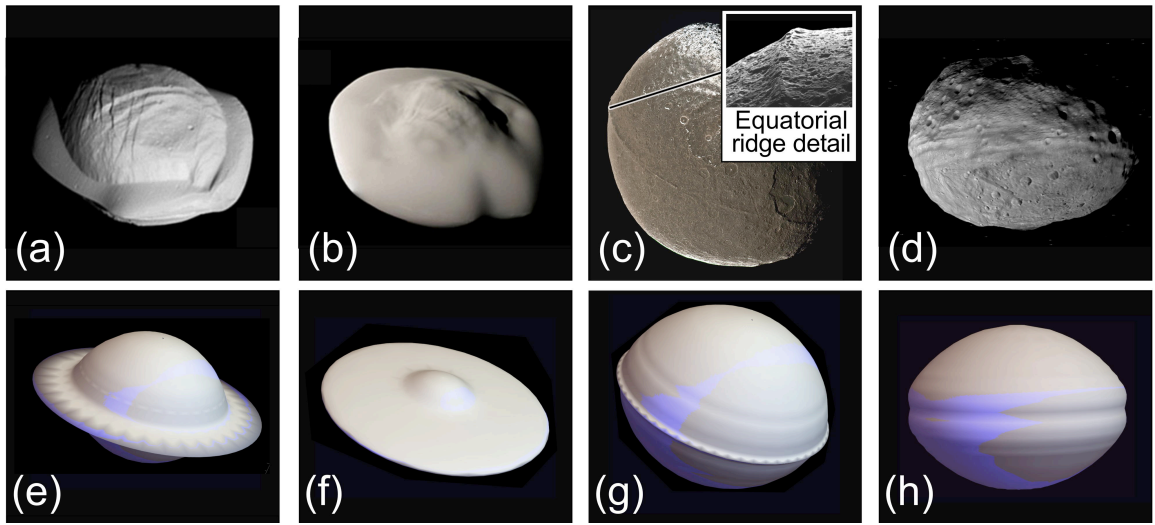

*Figure 1 – (a)-(d) Saturn's moons Pan, Atlas and Iapetus, and the asteroid Vesta (credit NASA/JPL-Caltech/SSI) compared with simulations (e)-(h) of morphology expected for freely flowing grains on a spheroidal mass influenced by a thin, weakly gravitational plane. Simulation parameters (defined in text): (e) $\omega = 5.15$, $R_{excised} = 4R_{moon}$, $R_{polar} = 0.85R_{moon}$, $\rho_{plane} = \rho_{moon}$; (f) same as panel (e) but with $\rho_{plane} = 1.2\rho_{moon}$, (g) $\omega = 8$, $R_{excised} = 3.2\,R_{moon}$, $\rho_{plane} = \rho_{moon}$; (h) $\omega = 25$, $R_{excised} = 2.5\,R_{moon}$, $\rho_{plane} = \rho_{moon}$. Geometry of simulations is shown in Fig. 2(c), where plane thickness is 2% of spheroid's major axis and plane outer radius is 10 times the radius of the moon in these panels.*

The origins of these features are unknown. Theories for the formation of equatorial ridges include collisions between matched pairs of bodies [6], viscous deformation of ring material [7], and accretion of material from a flat ring onto a central satellite [8,9]. No existing theory predicts equatorial troughs. We show here that troughs as well as ridges arise naturally if freely flowing grains follow equipotentials of a massive spheroid in a massive plane, as shown in Fig. 1(e)-(h). This possibility does not appear to have been examined previously.

The binary collision theory produces veridical equatorial ridges [6], but requires that the two bodies have nearly identical masses, and deform rather than fragment. Additionally, constraints on the collision impact parameter are stringent, both so that the bodies aren't torn apart by centrifugal forces and because any nonzero impact parameter would produce rotation perpendicular to the moons' equators – opposite of their current orientations. Moreover Pan and Atlas, and possibly Pandora and Daphnis as well, are quite similar in shape, making for an impressive coincidence of remarkable events.

The theories of ridge formation from ring material, on the other hand, involve highly probable mechanisms such as viscous spreading early in a moon's evolution [7] and deposition of ring material onto a satellite later [8,9]. The deposition model is supported by data from close fly-bys indicating that Pan, Atlas and other smaller moons are likely composed of a high density core surrounded by much lower density accreted material [10]. To date, however, deposition models have not reproduced the curious shapes shown in Fig. 1.

## Gravitational equipotentials:

Both types of theory are based on the presumption that equatorial features require a secondary morphogenetic process – collision, viscous spreading, or accretion – because equatorial features are not predicted from simple gravitational and inertial considerations. That is, as shown in cross section in Fig. 2(a), equipotentials of a rotating spheroid are themselves nearly spheroidal (for example in the blue cross section shown), and lack equatorial ridges or grooves. On the other hand, equipotentials for the same rotating spheroid embedded in a thin stationary planar disk can exhibit <u>both</u> a ridge (red) <u>and</u> a groove (green) shown in Fig. 2(b), and enlarged in Fig. 2(d). This means that either a ridge or a groove could form depending on the history of material deposition (discussed shortly).

All equipotentials shown are based on a solution to Laplace's equation for a uniform massive ellipsoid that has been known since 1840 [11]. Details of the solution appear elsewhere [12], and we provide an annotated numerical code for its analysis in Supplementary Materials accompanying this article. The essence of the solution is to break the gravitational geopotential into three parts: a net mass and a quadrupole term that define gravity outside of an ellipsoid of uniform density, and a centrifugal term that accounts for steady rotation. All terms are 3D, but we take the ellipsoid to have azimuthal symmetry – hence we show 2D cross sections in Fig. 2(a)-(b) and elsewhere. Surfaces of revolution are included to aid visualization in Figs. 1 & 5.

In this work we calculate equipotentials of a central spheroid embedded in a planar mass using the geometry sketched in Fig. 2(c), and we determine shapes of resulting surfaces assuming that deposited mass flows freely with minimal friction. For generality, our analysis defines the central mass rotating at constant angular speed $\omega$, with polar radius $R_{polar}$ and equatorial radius $R_{moon}$. For simplicity though, unless otherwise specified simulations are for a spherical body ($R_{polar} = R_{moon}$). We superimpose on the central sphere a planar mass that pierces the moon through its equator. It is expected [8] that mass near a moon will be accreted or expelled over time, so as shown in the figure we excise mass from the plane that lies within a radius $R_{excised}$ surrounding the moon.

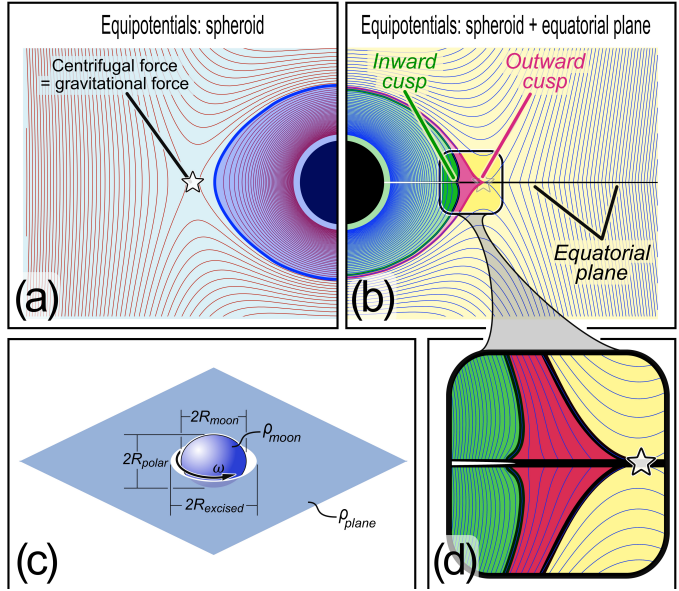

*Figure 2 –Gravitational equipotentials. (a) Equipotentials of central sphere (black) rotating with speed $\omega = 15$. (b) Equipotentials of the same sphere embedded in a plane whose density is half that of the central sphere, using the geometry shown in (c). (d) Enlargement of region near equator, highlighting both an inward and outward cusp. Equipotentials in (b) and (d) are for outer diameter of the plane $10R_{moon}$, plane thickness $0.02R_{moon}$, and where a radius $R_{excised} = 2.8R_{moon}$ of mass is removed from the plane as described in text.*

We define both the plane and the excised mass using the same 1840 ellipsoidal model as for the central spheroid, but we approximate a planar geometry by using very thin spheroids, with polar radii $0.01R_{moon}$ (i.e. planar thickness $0.02R_{moon}$). The plane has radius $R_{plane}$ much larger than $R_{moon}$: the value of $R_{plane}$ will be specified in example simulations, but is at least $10R_{moon}$ in all cases. The density of the plane, $\rho_{plane}$, will also be specified: we show results for densities ranging from $\rho_{plane} = 1.2\rho_{moon}$ to $\rho_{plane} = 10^{-4}\rho_{moon}$. In all cases, both the plane and the excised mass have zero angular speed. We take advantage of the fact that Laplace's equation is linear, so that we can obtain the gravitational potential of both the moon and plane by superimposing solutions for each. In order to excise mass from the plane, we simply define the density of the excised region to be negative: $\rho_{excised} = -\rho_{plane}$. The excised region overlaps the plane, so total mass is always non-negative, and varies smoothly from zero at the center of the satellite to $\rho_{plane}$ at the edge of the excised region. Mass is black, and excised region is white in Figs. 2 & 3.

To summarize, we calculate the gravitational potential, $U$, as the sum of three parts:

$$U = U_{moon} + U_{plane} + U_{excised}, \qquad [1]$$

where $U_{moon}$ is the potential for a spinning spheroidal moon with density $\rho_{moon}$, $U_{plane}$ is the potential of a plane defined to be a stationary spheroid with density $\rho_{plane}$ and thickness 1% that of the moon's equatorial diameter, and $U_{excised}$ is the potential of an excised mass, defined to be a stationary spheroid with negative density, $-\rho_{plane}$, thickness the same as the plane and radius $R_{excised}$ that we will vary. All three bodies are concentric as shown in Fig. 2(c). Each of the three parts is defined by the 1840 solution in ellipsoidal coordinates; conversions to and from Cartesian coordinates are included in Supplementary Materials.

The gravitational potential of a single spheroid has been calculated previously [13], but compares unfavorably with shapes of Saturn's moons. The equipotentials shown in Fig. 2(a) are typical, and use $\omega = 15$. Units are nondimensional and chosen for computational convenience, with a gravitational constant $G = 10^4$, and $\rho_{moon} = \frac{3}{4\pi}$, giving the moon unit mass (dimensional comparison for Saturn's moons below).

To examine the effect of a surrounding planar mass, we begin with a first, simplest, example: we consider a central sphere $R_{polar} = R_{moon} = 1$, so that gravitational acceleration at radius $r$ is just $a_{grav} = \frac{-G\,M_{moon}}{r^2}$. This competes against a centrifugal acceleration of $a_{cent} = \omega^2 r$, so $-a_{grav} = a_{cent}$ at the location indicated by the white star in Fig. 2(a).

The same solution superimposed on a flattened spheroidal plane as we have described is shown in Fig. 2(b). Parameters for this example are defined in the figure caption; in short, all parameters for the central spheroidal moon are as in Fig. 2(a), and the surrounding plane has density half that of the moon: $\rho_{plane} = \frac{1}{2}\rho_{moon}$. The outer radius of the plane is 10 times that of the moon: $R_{plane} = 10 R_{moon}$, and the excised radius is chosen to be $R_{excised} = 2.8 R_{moon}$ for reasons that we will explain. We discuss next effects of varying each of these parameters; for the time being, we note that both inward and outward cusps arise naturally as equal gravitational potentials of a central moon embedded in a plane as can be seen from the enlargement of Fig. 2(d).

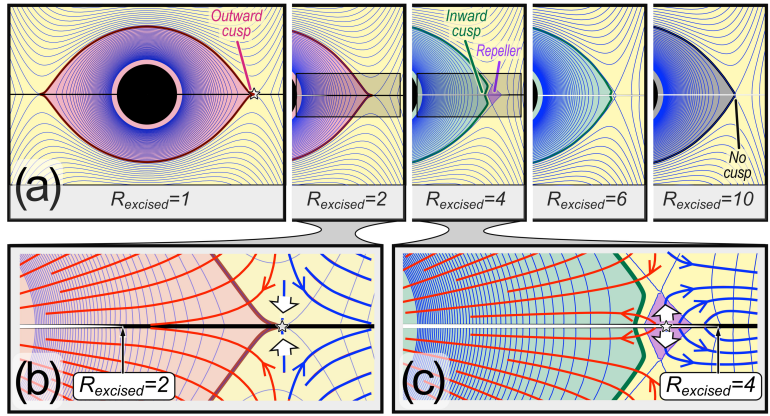

*Figure 3 – Mechanism of bifurcation between outward and inward cusp. (a) Progression of cusp shapes as distance, $R_{excised}$, of plane mass from central sphere (black, with radius $R_{polar} = 1$) is increased. (b) Enlargement of equipotentials (thin curves) and streamlines (thick curves) for $R_{excised} = 2$. The satellite has an outward cusp (pink, bounded by dark red), and all streamlines move toward the equatorial line in black. Red streamlines bring mass toward the satellite and flow toward a single saddle point is indicated by the large white arrows; blue streamlines bring mass away from the satellite. (c) Equipotentials and streamlines for $R_{excised} = 4$: the white arrows have reversed direction, and all mass within the violet region moves outward, ultimately reaching either the satellite (now green, with an inward cusp) or the plane (black). In all panels, $R_{plane} = 10^2 R_{polar}$, and the plane has thickness $0.02 R_{polar}$ as before.*

Two related remarks concerning the equipotentials shown must be stressed before continuing. First, it is important to make clear the meaning of equipotentials. An equipotential is by no means the only possible shape of a satellite: equipotentials are surfaces free of tangential stress, so they are the shapes that would be adopted if surficial material (e.g. freely flowing grains) fully relaxes in response to gravitational and centrifugal forces. Future investigations of accretion history as well as visco-elastic effects, are clearly merited: the intent here is to present shapes that would develop in the freely-flowing limit.  Fig. 2 shows multiple equipotentials; the choice of which is adopted depends on the amount of mass present.  The equipotentials identified in green and red in Fig. 2 are chosen to define surfaces with the most pronounced inward or outward cusp respectively.  If more material were deposited onto the green surface and allowed to relax freely, the cusp would diminish its curvature, whereas if material were added to the red surface and again allowed to relax freely, it would be stripped away by centrifugal acceleration.

Second, the solutions entirely neglect gravitational interactions between the body and freely flowing material being accreted.  These interactions change the gravitational shape of the body, which in turn changes the shape of the equipotentials, but this fully nonlinear problem is non-integrable, and a complete solution is neither unique nor generally attainable.  We do discuss gravitational influences of a parent body (here, Saturn) at the conclusion of this paper, but we emphasize that the satellite shapes that we display – either with or without Saturn's gravity – are idealized.

With this in mind, to understand the mechanism underlying cusp formation, we observe that cusps emerge as the plane approaches the central spheroid.  That is, as $R_{excised} \to R_{plane}$ the effect of the plane vanishes and we recover the cusp-free solution shown in Fig. 2(a), while as $R_{excised}$ diminishes, the influence of the plane on the moon's gravity grows.  So we can examine the effect of the plane by varying $R_{excised}$ between $R_{moon}$ and $R_{plane}$.  This is shown in Fig. 3(a) for the representative case $\omega = 15$ (again, variations in $\omega$ will be considered).  Parameter values used in Fig. 3 are defined in the caption; we have found in multiple trials that the progression shown is quite typical for a wide range of parameter choices.

Fig. 3(a) and the enlargement of panel b show that a nearby concentric plane, $R_{excised} \leq 2$, produces only an outward cusp (seen also in Fig's 1(e)-(g)).  For a more distant plane, $R_{excised} \geq 3$, the outward cusp bifurcates to form an inward cusp and a repeller, highlighted in the enlargement of panel c.  Any mass within the region labeled "Repeller" will gravitate either toward the moon or toward the surrounding plane, along the streamlines plotted.  Streamlines are defined in the usual way from the potential $U = U_{moon} + U_{plane} + U_{excised}$ defined earlier.  As discussed next, the bifurcation of the outward cusp into an inward cusp combined with a repelling region is more intricate than suggested by Fig. 3(a): evidence of this intricacy is seen in Fig. 2(b), where we see in a simulation using the same parameters as in Fig. 3 that both inward and outward cusps are present between $R_{excised} = 2$ and $R_{excised} = 3$.

Putting details of the bifurcation aside for a moment, the coarse-grained outcome is that if a massive plane is near a satellite, it will tend to produce an outward cusp, and as the planar mass moves away from the moon, the outward cusp will give way to an inward one.  As mass recedes still further from the moon, its effect diminishes until equipotentials are indistinguishable from those with no plane at all, shown to the right of Fig. 3(a) at $R_{excised} = 10 R_{moon}$.

The mechanism underlying cusp formation can be further exposed by examination of the enlargements of Fig. 3(b)-(c).  If a massive plane is near the moon, shown in Fig. 3(b), all nearby mass is drawn equatorward, as the streamlines indicate.  This either brings nearby mass toward the moon (red arrows) producing an outward cusp, or brings it away (blue arrows), to augment the surrounding plane.  When the plane recedes past the unstable point indicated by the star, a bifurcation reverses mass flow near the starred point.  This can be seen by comparing the flow directions shown by the white arrows in Fig's 3(b) and 3(c).  This bifurcation arises when the planar mass moves from inside the region identified by the red equipotential in Fig's 3(a)-(b) (drawing mass inward, toward the moon) to outside that region (drawing mass outward, toward the surrounding plane).  Between these two extremes, both states can co-exist, as shown in Fig. 2(b).

Dynamically speaking, the violet region surrounds what would be the Lagrange point, L1, if there were a single orbiting mass rather than mass within a plane. In our case, for small $R_{excised}$, L1 is a saddle, while for larger $R_{excised}$ the saddle splits in two and L1 becomes an unstable node. This means that an unstable torus devoid of mass can be expected to surround a moon with an inward cusp embedded in a planetary ring (cf. Vesta). We will see shortly that a stable torus is also possible.

In Fig. 3 we held $\omega$ fixed and varied $R_{excised}$; we next consider effects of varying both $\omega$ and $R_{excised}$. This results in Fig. 4(a), where we plot states near the equator for different values of these parameters. As with Fig. 3, this plot is coarse-grained, so it does not detail finer features such as the narrow bistable region that persists at least up to $\omega = 15$ (as shown in Fig. 2(b)). Additionally, we emphasize that the equipotentials shown are obtained through a sequence of computational steps including transforms of coordinates both to and from elliptical coordinates and implicit solutions of a superposition of states for the central ellipsoid and two nearly planar ellipsoids. These calculations (see attached code) necessarily have limited accuracy, and as a consequence we report here only unambiguous large-scale features, and disregard smaller-scale intricacies that may or may not be reliable. Our criteria for whether a feature is reliable are whether it changes smoothly with parametric variations, and whether it is larger than observed computational artefacts. For example, the violet repeller shown in Fig. 3(c) grows with both $\omega$ and $R_{excised}$, and is seen for other choices of $\rho_{plane}$. On the other hand, slopes of computational equipotentials near the equator change direction erratically, and so features that never grow above a few percent of $R_{polar}$ are not reported here.

Bearing these caveats in mind, Fig. 4 reveals the presence of <u>five distinct states</u> of freely flowing regolith subject to gravitational equipotentials for satellites embedded in a massive plane. For small $R_{excised}$, shown in Fig 4(b) for $R_{excised} = 2$, $\omega = 2.5$, an outward cusp is present for any satellite rotational speed, $\omega$. Larger $\omega$ values causes material further from the satellite to be centrifuged away: this both reduces the equatorial radius of the equilibrium point at which $-a_{grav} = a_{cent}$ and shortens outward cusps. So elongated outward cusps can be expected at small $R_{excised}$ and small $\omega$, as shown in Fig. 4(b).

As $R_{excised}$ grows, an inward cusp emerges, shown in Fig. 4(c) for $R_{excised} = 5$, $\omega = 10$. As we have mentioned, along with this change in cusp direction, a region appears between the satellite and the plane that repels all mass. Between the outward and inward cusp, a third state is possible that supports both cusps. We have already commented briefly on this bistable state, which is shown in Fig. 2(d) for $R_{excised} = 2.8$, $\omega = 15$. A fourth state arises when the planar mass is far enough away from the central satellite that the inward cusp disappears: in this case, the classical shape shown in Fig. 2(a) is recovered.

Finally, a new, fifth, state is predicted by this model, shown in Fig. 4(d): for small rotational speeds and for a massive plane around $R_{excised} = 8R_{moon}$ away, an attracting region appears through a saddle-node bifurcation. This state is shown in Fig. 4(d). We note that the repelling and attracting tori should be present for either an inward or an outward cusp, but are unlikely to be observable for an outward cusp since they will be inside the satellite's surface. Nevertheless the repelling torus may represent an area of weakness within such a satellite that could cause equatorial mass to break away from its main body. More likely, the attracting state could spontaneously produce a torus around a satellite, and mass injected nearby could become entrained in such a satellite's orbit.

We reiterate that all of these states form for free surface flow under linear and steady-state conditions. History dependence, self-gravitation, and viscoelastic behavior of satellite material are neglected. Processes that generated cusps on Saturn's moons doubtless occurred long ago, and so $R_{excised}$ and $\omega$ values in Fig. 4 may be useful for evaluating conditions when cusps formed, but are not representative of current conditions.

Nevertheless, we can orient ourselves on Fig. 4 using current data, as indicated by the "Pan today" marking in panel (a). We have mentioned that units in that panel are dimensionless, and are obtained for a central sphere of unit radius and mass, and a time scale set by choosing the gravitational constant (we use $G_{simulated} = 10^4$ for plotting convenience). For the purpose of orientation, we consider Saturn's moon Pan as an example. Pan lies within the Encke gap, with width 325 km: about 20 times Pan's radius, so $R_{excised}$ is at least 20 currently. Pan has density $\rho = 420$ kg/m³ and is tidally locked with frequency $\omega = 2\cdot 10^{-5}$ sec⁻¹, which we can relate to the ordinate in Fig. 4 by making use of the dimensionless group, $\mathcal{G} = \frac{G\cdot\rho}{\omega^2} = \frac{6.7\cdot 10^{-11}\cdot 420}{(2\cdot 10^{-5})^2} = 70$. Applying the same dimensionless group to our simulations, $\mathcal{G} = 70 = \frac{G_{simulated}\cdot\rho_{simulated}}{\omega^2_{simulated}} = \frac{10^4\cdot 0.24}{\omega^2_{simulated}}$, so the corresponding simulated frequency is $\omega_{simulated} \sim 6$, at the location shown in Fig. 4(a).

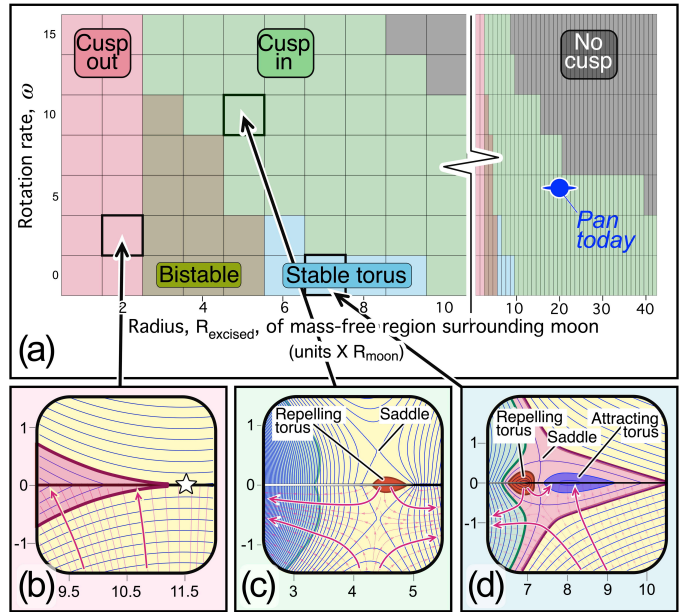

*Figure 4 – Phase diagram of equatorial features of freely flowing material. (a) Five distinct states identified as rotation rate $\omega$ and central excised mass $R_{excised}$ are varied. Current parameters for the moon Pan are shown for illustrative purposes. Evidently Pan's outward cusps would not form today, and would require disk material much closer than currently seen. (b)-(d) Enlargements of equipotentials and streamlines in cusp regions. Horizontal axes align with disk; axis units agree with those defined earlier, i.e. the central spheroid has radius $R_{polar} = 1$, and the origin is at center of the spheroid. (b) For nearby plane (small $R_{excised}$), all mass gravitates equatorward, producing an outward cusp. (c) For more distant plane, the bifurcation shown in Fig. 3 reverses the flow near a repelling torus, producing an inward cusp. (d) For moderate $R_{excised}$ and small $\omega$, a novel attracting torus emerges, bracketed between an inward and an outward cusp; a 3D rendition of the attracting torus appears in Fig. 5(a). In all panels, $R_{plane} = 10^2 R_{polar}$, and the plane again has thickness $0.02 R_{polar}$.*

To examine the effect of a surrounding planar mass, we begin with a first, simplest, example: we consider a central sphere $R_{polar} = R_{moon} = 1$, so that gravitational acceleration at radius $r$ is just $a_{grav} =$ $\frac{-G\,M_{moon}}{r^2}$. This competes against a centrifugal acceleration of $a_{cent} = \omega^2 r$, so $-a_{grav} = a_{cent}$ at the location indicated by the white star in Fig. 2(a).

In Fig. 5(a), we also plot the stable torus' location for the conditions shown in Fig. 4(d): $R_{excised} = 7$, $\omega = 0$, and $\rho_{plane} = \frac{1}{2}\rho_{moon}$, $R_{plane} = 100 R_{polar}$ (the parameters used in the other cases shown in Fig's 3 & 4). The image in Fig. 5(a) displays the outside of the attracting torus: freely flowing material will gravitate toward a small region at its center; moreover we emphasize that tidal perturbations and impacts can remove this material, and the torus will become unstable as the planar mass recedes. Nevertheless, straightforward analysis of equipotentials indicates that a stable torus ought to be present in some satellites within a planetary ring. Interestingly, this torus appears at small – even zero – angular speed $\omega$, thus material in the torus can rest stably without orbital velocity!

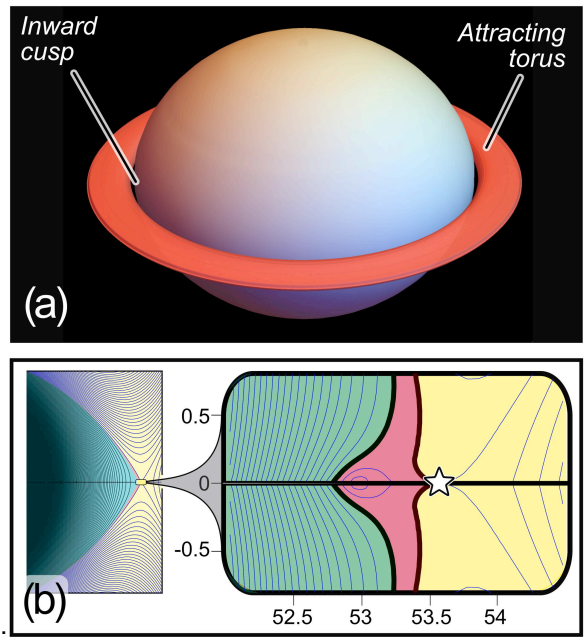

*Figure 5 – Extremes in cusp behavior. (a) 3D reconstruction of attracting torus from Fig. 4(d). The torus shown is at the outer extreme of the attractive region, and mass would gravitate toward a central torus within. (b) Equipotential cusps for very low planar density: $\rho_{plane} = 10^{-4} \rho_{moon}$; axis units as in Fig. 4. Left panel shows equipotentials to right of the attracting body; right panel shows enlargement, including both inward (green) and outward (red) cusps, as well as equal force point identified by star. Parameters are $R_{plane} = 10^3 R_{polar}$, $R_{excised} = 53 R_{polar}$, $\omega = 0.25$, and central mass is a sphere and the plane again has thickness $0.02 R_{polar}$.*

Two final, practical, issues remain. First, arguably the mass of Saturn's disk material could seem to be too small to produce significant gravitational effects on its moons, and second, effects of Saturn's gravity (mentioned earlier) remain to be discussed. We consider these two issues next.

## Saturn's disk mass:

With respect to the mass of Saturn's disk material, two facts suggest that this mass could be sufficient to produce the satellite shapes that we have displayed. First, the density of Saturn's rings was almost certainly higher during the evolution of its moons than it is today [8]. Indeed, prior work [14] has estimated the ratio of the mass of rings to moons to have been as high as 1:3. By the same token, satellites (e.g. Iapetus) or asteroids (e.g. Vesta) that are currently far from disk material may have been embedded in such material earlier during their formation. Second, we can quantify how large the density of a surrounding plane must be to produce equatorial cusps, and we will next show that cusps arise even for extremely small planar densities.

To evaluate effects of planar density, $\rho_{plane}$, we perform an additional set of simulations holding $\rho_{moon}$ fixed and reducing $\rho_{plane}$ by successive orders of magnitude to determine the point at which cusps vanish. Specifically, at each value of $\rho_{plane}$, ranging from $\rho_{plane} = \rho_{moon}$ to $\rho_{plane} = 10^{-5}\rho_{moon}$, we manually adjust $R_{excised}$ and $\omega$ to determine whether cusps are present. This is easily done by making use of the facts first that cusps are stabilized by reducing $\omega$ (which reduces centrifugal stripping of material from the cusp) and second that cusps tend toward higher $R_{excised}$ at smaller $\omega$ (cf. Fig. 4). So by examining equipotentials near the starred equilibrium point while incrementally reducing $\omega$ and increasing $R_{excised}$, cusps are readily identified.

From this exercise, we find that cusps persist at least down to $\rho_{plane} = 10^{-4}\rho_{moon}$: we show both inward and outward cusps in Fig. 5(b) for this case: these appear with the plane held at a distance, $R_{excised}$, over 50 times the central mass' radius, and for very small rotation rate. Even smaller cusps may be present at lower planar densities, but they are at most a few percent of $R_{polar}$ in size, and as we have mentioned this is the same size as computational noise and so are disregarded here. Ultimately we conclude that even planar disk densities 1/10,000[th] that of a central spheroid – and at a distance 50 times the radius of the central mass – are sufficient to produce equatorial cusps in gravitational equipotentials. By comparison, the moons Pan and Atlas have densities estimated [9] to be 410-450 kg/m$^3$, while recent calculations [15] indicate that the densities of Saturn's rings range from zero up to 60 kg/m$^3$: a difference of up to about 1/10.

It remains uncertain how dense the rings once were, but comparison with the 1/10,000 figure that produces cusps suggests that sufficient ring density may be present, even currently, to affect satellite geomorphology. Moreover as we have mentioned, the observation that a interplanetary body is currently far from a planar mass does not imply that the body was not influenced by such a mass earlier in its history. Since it is not possible to determine complete histories of these bodies in retrospect, we propose that their current morphologies can be used to provide evidence of that history *a posteriori.* That is, on mathematical grounds, we find that circumferential ridges and grooves on celestial bodies may be indications that the bodies at one time were embedded in a planar mass and covered with freely flowing regolith.

## Saturn's gravity:

This brings us to the final practical issue: effects of Saturn's gravity. Here we note that Saturn's gravity at the current locations of Pan and Atlas is 3 orders of magnitude stronger than their surface gravities. From that perspective, the relevance of analyzing gravitational equipotentials needs justification.

Pan and Atlas are currently about $R_{orbital} = 1.3 \cdot 10^8$ m from Saturn, and are maintained in orbit by outward centrifugal acceleration. Explicitly, at that distance, the net inward acceleration acting on a moon in a circular orbit is:

$$a_{moon} = \frac{G\,M_{Saturn}}{R_{orbital}^2} - \frac{v^2}{R_{orbital}}, \qquad [2]$$

where $G$ is the universal gravitational constant, $M_{Saturn}$ is Saturn's mass, and $v$ is the moon's orbital speed, all of which are known. At equilibrium, $a_{moon}$ at a moon's center of mass is zero, while the difference in acceleration from the near to the far side of a moon with radius $r_{moon}$ is given by:

$$\Delta a_{moon} = \left(-\frac{2G\,M_{Saturn}}{R_{orbital}^3} + \frac{v^2}{R_{orbital}^2}\right) 2r_{moon}. \qquad [3]$$

This is the maximum acceleration responsible for tides [16], and is readily evaluated. For the four examples shown in Fig. 1, from right to left, Vesta is not in Saturn's orbit, and both $a_{Vesta}$ and $\Delta a_{Vesta}$ are zero. So if Vesta were in a massive plane far from a central body earlier in its history (which is not known), the equipotentials presented earlier would hold. Iapetus is very massive and is currently far from Saturn, so $\Delta a_{Iapetus}$ is small: under $10^{-4}$ times its surface gravity, and again there is justification for neglecting the mass of a central body. Pan and Atlas, on the other hand, are both less massive and closer to Saturn: $\Delta a_{Pan}$ and $\Delta a_{Atlas}$ are respectively about 1/4 and 1/5 of the moon's surface gravity.

Focusing on these two moons that are most strongly influenced by Saturn's gravity, Atlas' rotation rate is not known, but Pan is in synchronous rotation with the same side always facing Saturn, so tidal forcing on its surface is static and its perturbation to the equipotential due to Saturn is bounded by the following term:

$$U_{tidal} = -\int \Delta a_{moon}\ d\vec{r}_{moon}, \qquad [4]$$

where $\vec{r}_{moon}$ points toward Saturn. Adding this to the potential, Eq. [1] provides us with an idealized potential from a perturbation due to Saturn's gravity.

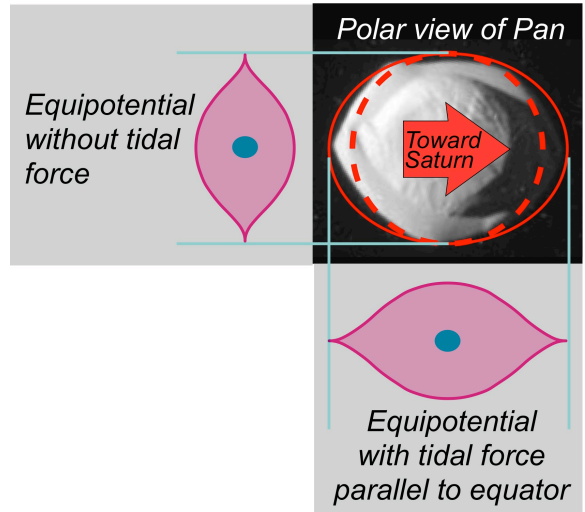

*Figure 6 – Pan and orthographic depictions of distortion of gravitational equipotentials by tidal force from distant planet. Upper right shows polar view of Pan with superimposed circle (dashed curve) and ellipse (solid). The ellipse's semi-minor axis is obtained from the maximum radius of potential [1]; cross-section shown in upper left. The ellipse's semi-major axis is obtained by adding the potential [4] with $\vec{r}_{moon}$ aligned with the horizontal, with a convenient magnitude of the maximum tidal acceleration. Other magnitudes are included in the Appendix. Pan credit: NASA/JPG-Caltech/SSI.*

In Fig. 6, we plot a polar photograph of Pan alongside orthographic views of the deformation that one would naïvely obtain from Eq. [4]. The superimposed dashed line in the polar view is a simple circular outline, as would be expected without a perturbing gravitational potential. The solid line shows an ellipse with semi-minor axis of the circle and semi-major axis obtained by adding the perturbation of Eq. [4]. The arrow indicates Pan's orientation with respect to Saturn, and the orthographic views give respective cross-sections that one would expect perpendicular to (i.e. uninfluenced by) Saturn's field, and parallel to Saturn's gravity.

This is again only an idealized analysis and is far simpler than the complete problem – for example both the temporal reshaping due to long periods of tidal damping and important nonlinear interactions are overlooked. Likewise, continuing the spirit of the rest of our analysis, we neglect changes to equipotentials due to redistribution of mass, here from the orbital direction to the Saturn-moon direction. The merit of Eqs. [1] and [4] is that they permit us to calculate equipotentials, which give the lowest order stress-free state of a satellite's surface that could be expected to be adopted by a linear analysis of individual and sedately settling particles.

## Conclusion:

It is well recognized that Saturn's moons affect its rings [17], for example producing wakes, gaps and resonances that dominate ring morphology [18]. Effects that rings have on moons are less well studied [8], and gravitational potentials of satellites are assumed to be insensitive to influences of rings. Contrary to this assumption, we have found that distinctive features of Saturn's moons, including equatorial cusps and even a predicted stable circumferential torus, arise naturally by including the influence of rings in the simplest possible gravitational model of a central satellite surrounded by a massive equatorial plane. Remarkably, equatorial cusps endure even for distant and low mass rings. It is important to emphasize that these results are purely linear: like the effect that moons have on rings, we neglect nonlinear feedback between rings and moons. It is likely – especially since ring densities several orders of magnitude smaller than moon density have significant gravitational effects – that this feedback is important, and indeed the model presented here does not yet consider how accreted mass affects a satellite's gravitational potential. A fully nonlinear model continues to be an important challenge. Likewise, as we have mentioned, results here apply in the limit of entirely freely flowing grains: effects of accretion history and of friction, viscosity, and elasticity also remain for future investigation.

## Acknowledgments:

*This material is based on support from NSF CBET, award no. 1804286.*

# *Appendix*

## Effects of parent gravitational body

In this section we briefly present results of varying the strength and location of an external, distant, gravitational body. These results are tangential to Fig. 6, which concerns perturbations to equipotentials due to a parent gravitational body (Saturn in the case of Pan and Atlas). Only the lowest order terms of the tidal potential are considered, and the parent body is assumed to be distant enough that its gravitational fieldlines can be treated as parallel. The field then drops off as $1/R_{orbital}^2$, where $R_{orbital}$ is the distance between centers of mass of the parent and the moon, as shown in Eq. [2] in the main text.

The net force on a particle on the surface of a moon then depends on relative magnitude and direction between the parent body's gravitational field (Eq. [2]) and the gradient of the moon's equipotentials (Eq. [1]). That is, if the parent body's net force vector aligns with the vector of the moon embedded in a plane, one will get different perturbed equipotentials than if the two vectors are in different directions.

In Fig. A1, we plot the two extremes in direction of the parent body with respect to the moon and surrounding plane. First, in panel (a) we plot equipotentials for the case in which the parent body is in the surrounding plane, so tides are parallel to the plane (denoted $tide_{\parallel}$), and second in panel (b) we plot the case in which the parent body is located perpendicular to the surrounding plane (denoted $tide_{\perp}$). In both cases, we assume that the moon is in synchronous rotation, so that its orientation toward the parent body is unchanging, and in both sets of plots we indicate the tidal direction (toward Saturn) by red arrows. "No tide" applies to a free satellite embedded in a gravitational plane, and tidal amplitudes are expressed as fractions of a moon's gravity at the equator of the unperturbed satellite (i.e the $tide = 0$ case). So for example $tide_{\parallel(\perp)} = 14\%$ means that an acceleration with magnitude of 14% of a satellite's gravity at the unperturbed equator is applied in direction parallel ($\parallel$) or perpendicular ($\perp$) to the surrounding plane. In all cases, we take the moon to rotate slowly ($\omega = 5.15$), and to be embedded in a plane that reaches the moon ($R_{excised} = R_{moon}$). As shown in Fig. 4(a) of the main text, these values produce a single outward cusp. The plane's density and size is as in the left panel of Fig. 3(a) of the main text: $\rho_{plane} = \frac{1}{2}\rho_{moon}$, $R_{plane} = 10^2 R_{polar}$, and plane thickness $0.02 R_{polar}$.

Fig. A1(a) shows that the perturbation by a parent body (Eq. [2]) produces only a quantitative change in equipotentials, simply elongating the moon's ravioli shape toward the parent body. Fig. 6 in the main text uses $tide_{\parallel} = 8\%$: a value chosen to roughly correspond to Pan's apparent distortion.

Fig. A1(b) by contrast shows that if the two vectors are perpendicular, qualitative bifurcations arise. Without a parent body ("No tide"), the usual Lagrange points are supplemented with two saddle points, denoted S1 and S2.

We make two orientational remarks. First, the plots of Fig. A1 are all 2D cross-sectional views of a 3D problem, so it should be borne in mind that elliptic points could equally be cross sections of deformed ellipsoids or of tori: the choice depends on details outside of the cross-sectional plane.

Second, the saddle points, S1 and S2, are in the surrounding massive plane, and Fig. A1(b) considers the parent body to be perpendicular to that plane. If the parent is in the plane, one obtains Fig. A1(a). Thus the results of Fig. A1(b) are likely academic and would not apply to Saturn and its disk in its current state. Nevertheless, Saturn's disk inclination with respect to the ecliptic is 27°, and so the same analysis as presented here could be used to describe the Sun as a parent gravitational body perturbing equipotentials surrounding Saturn. Indeed, Uranus' disks are nearly perpendicular to the ecliptic, so it is possible that equipotentials of Uranus' irregular (Nesvorný, 2007) moons could be deformed by the sun as suggested by Fig. A1(b). In short, scenarios in which the orientations of moons, disks, and external bodies are misaligned may be speculative, but are not implausible.

Bearing this in mind, as the parent body's gravitation grows from zero to to about 14% of the moon's unperturbed equatorial magnitude, Fig. A1(b) shows that an elliptic Lagrange point approaches the moon. This point is in a plane connecting the moon and parent, but is perpendicular to the plane of the disk surrounding the moon. We denote this point L1 to agree with the usual nomenclature in the absence of a disk, but note that in the presence of a massive disk L1 is elliptic, rather than hyperbolic as it would otherwise be.

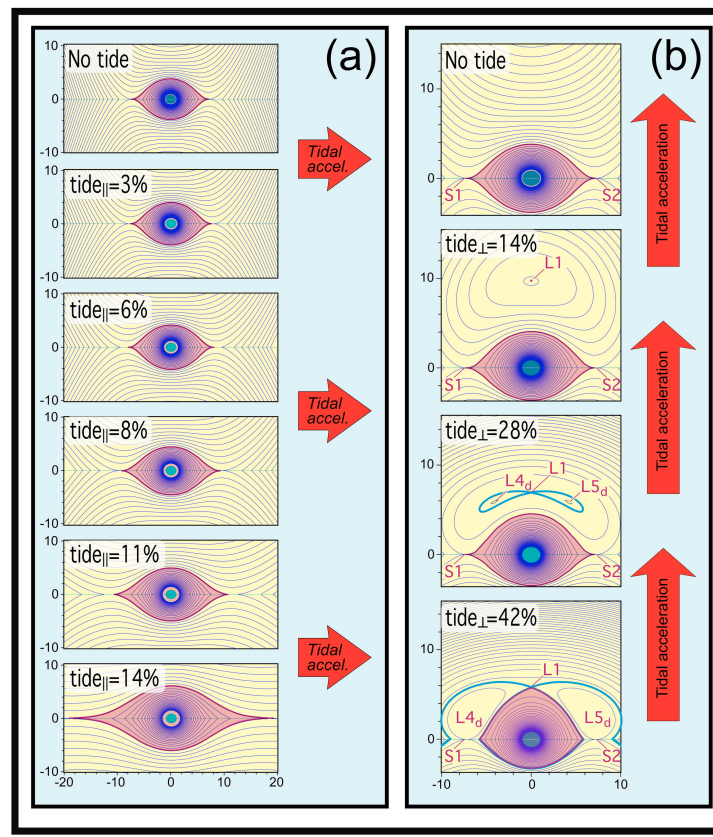

*Figure A1 – <u>Effects of gravitation by parent body</u>: Cross sections of gravitational equipotentials using Eqs. [1] plus Eq. [2] in the main text to include effects a parent body on a rotating moon in a massive disk. Parameters used are defined in the text. (a) Equipotentials where the parent body is in the plane defining the moon's rotation and its surrounding disk (as is Pan). (b) Equipotentials where the parent body is perpendicular to that plane. Lagrange points are denoted by L, and $S_1$ & $S_2$ are saddle points not present without effects of the planar mass. As described in the text to this Appendix, a parent body in the plane produces only elongation of the equatorial cusp, whereas a parent body perpendicular to the plane produces a pitchfork bifurcation (between $tide_\perp = 14\%$ and $28\%$) followed by a change of allegiance of the point L1 (between $tide_\perp = 28\%$ and $42\%$) as the parent body's gravity increases.*

As the parent's gravitation grows to about 28%, L1 returns to its hyperbolic state following a Hamiltonian pitchfork bifurcation. To conserve topological index, this bifurcation necessarily produces a pair of elliptic points, which we denote $L4_d$ and $L5_d$. The bifurcation is accompanied by a polar bulge in equipotentials surrounding the moon. Unlike the usual Lagrange points L4 and L5, the points $L4_d$ and $L5_d$ make no connection below (in the orientation shown) the moon, and to record this distinction we indicate the presence of a disk with the subscript "d". At larger parent gravitation yet, the point L1 changes allegiance (Shinbrot, 1996), and its homoclinic connection becomes heteroclinic, making new connections to S1 and S2. This state persists as the parent gravitation grows to dominate over the disk gravity. At this point L1 more closely approaches the moon and $L4_d$ and $L5_d$ recede to become the more usual L4 and L5.

In summary, we find that a parent body in the plane surrounding a moon (as with Saturn and its moons) produces only minor elongation in a moon's ravioli-shaped equipotentials, while a parent body located perpendicular to this plane (as perhaps with the Sun and Uranus) produces qualitative bifurcations in Lagrange points, and a noticeable polar bulge facing the parent body.

## Mathematica computer code

Ellipsoidal gravity definitions:

GM is a magnitude for the gravitational force, where M the mass interior to the ellipsoid,
a the equatorial radius,
b the polar semi-axis,
ee is measure of eccentricity: $ee^2 = a^2 - b^2$,
$\omega$ the rotation rate about the polar axis,
u is the 'distance' of the equipotential line from the origin,
$\beta$ is the parametric latitude,
RR is cylindrical radius,
ZZ is cylindrical z-coord.

Cylindrical coordinates RR and ZZ are obtained from ellipsoidal coordinates u and $\beta$ using:

RR = $\sqrt{u^2 + ee^2}$ Cos[$\beta$];
ZZ = u Sin[$\beta$];
z = $\frac{ee}{u}$

Conversion from cylindrical into ellipsoidal coordinates:

This is not needed for calculations here, but the solution is used to produce the functions beta[] and uu[] below. Note that there are 16 solutions accounting for all symmetries and antisymmetries; only 1 is needed.

In[34]:= 
```
ClearAll["Global`*"]
thetsfm = Quiet[Solve[{R1 == √(u² + ee²) Cos[β], Z1 == u Sin[β]}, {u, β}]];
```

Helper functions used in final solution:

In[36]:= $\text{beta}[RR\_, ZZ\_, EE\_] = \text{ArcCos}\left[\sqrt{\dfrac{EE^2 + RR^2 + ZZ^2 - \sqrt{EE^4 + RR^4 + ZZ^4 + 2\,EE^2\,ZZ^2 + 2\,RR^2\,ZZ^2 - 2\,EE^2\,RR^2}}{2\,EE^2}}\right];$

$\text{uu}[RR\_, ZZ\_, EE\_] = \dfrac{ZZ}{\text{Sin}[\text{beta}[RR, ZZ, EE]]};$

$\text{Quad}[z\_] = \dfrac{1}{2\,z^3}\left(\left(1 + \dfrac{3}{z^2}\right)\text{ArcTan}[z] - \dfrac{3}{z}\right);\ (*\ \text{function used for quadrupole}\ *)$

$\text{Um}[u\_, a\_, b\_, GM\_] = \dfrac{GM}{\sqrt{a^2 - b^2}}\,\text{ArcTan}\left[\dfrac{\sqrt{a^2 - b^2}}{u}\right];\ (*\ \text{mass term}\ *)$

$\text{Uq}[u\_, \beta\_, \omega\_, a\_, b\_] = \dfrac{\omega^2}{2}\,\dfrac{a^2\,b^3}{u^3}\,\dfrac{\text{Quad}\!\left[\frac{\sqrt{a^2-b^2}}{u}\right]}{\text{Quad}\!\left[\frac{\sqrt{a^2-b^2}}{b}\right]}\left(\text{Sin}[\beta]^2 - \dfrac{1}{3}\right);\ (*\ \text{quadrupole term}\ *)$

$\text{Ur}[u\_, \beta\_, a\_, b\_, \omega\_] = \dfrac{\omega^2}{2}\left(u^2 + a^2 - b^2\right)\text{Cos}[\beta]^2;\ (*\ = \dfrac{\omega^2}{2}R^2:\ \text{centrifugal term}\ *)$

$\text{Utotal}[u\_, \beta\_, GM\_, a\_, b\_, \omega\_] = \text{Um}[u, a, b, GM] + \text{Uq}[u, \beta, \omega, a, b] + \text{Ur}[u, \beta, a, b, \omega];$

Equipotentials for central ellipsoid combined with plane as described in Shinbrot "Gravitational influence of Saturn's rings on its moons":
Only one quadrant is plotted: this is associated with a choice of symmetric, antisymmetric solutions mentioned above. Other quadrants can easily be provided, but doing so is computationally redundant.

In[42]:= `SetOptions[Manipulator, Appearance → "Labeled"];` (* optional *)
`MaxOuterDisk = 100;` (* outer radius of plane, $R_{plane}$ *)
`gravfactor = 1/2;` (* factor by which density of plane < density of moon *)

(* following is used by Mathematica to allow changes in parameters with sliders *)
`Manipulate[`
  `theee = ` $\sqrt{\text{themajor}^2 - \text{theminor}^2}\,;$
  (* following is plot of planar mass: *)
  `p00 = Graphics[{Darker[Cyan], Opacity[0.9], {Disk[{0, 0}, {themajor, theminor}], Disk[{0, 0}, {MaxOuterDisk, diskthickness}]}},`

```
       White, Disk[{0, 0}, {negratio, diskthickness}]]}, PlotRange → {{-plotmin - plotradius, -plotmin}, {0, plotradius}}];

(* following is definition of sum of net mass, quadrupole mass and centrifugal acceleration: *)
SumFun[XX_, ZZ_, theGMFlat_, negratio_, themajor_, theminor_, theee_, theGM_, theomega_] =

  Utotal[uu[ √(XX² + 0²) , ZZ, √(MaxOuterDisk² - diskthickness²) ],
   beta[ √(XX² + 0²) , ZZ, √(MaxOuterDisk² - diskthickness²) ], theGMFlat, MaxOuterDisk, diskthickness, 0]

   - Utotal[uu[ √(XX² + 0²) , ZZ, √((negratio themajor)² - diskthickness²) ],
    beta[ √(XX² + 0²) , ZZ, √((negratio themajor)² - diskthickness²) ], theGMFlat / (MaxOuterDisk² / (negratio themajor)²)
    , negratio themajor, diskthickness, 0]

   + Utotal[uu[ √(XX² + 0²) , ZZ, theee], beta[ √(XX² + 0²) , ZZ, theee], theGM, themajor, theminor, theomega];

(* following is plot of equipotentials calculated above: *)
p01 = ContourPlot[SumFun[XX, ZZ, theGMFlat, negratio, themajor, theminor, theee, theGM, theomega] ,
  {XX, -plotmin - plotradius, -plotmin}, {ZZ, 0, plotradius}, Contours → Range[mincontour, maxcontour, numcontours]
  , ContourShading → None];

(* following displays both equipotentials and surrounding plane: *)
Show[p01, p00],

(* following are parameters that can be changed using sliders: *)
{{theGM, 10 000, "Moon Gravity"}, 0, 50 000},
{{theGMFlat, gravfactor theGM diskthickness (MaxOuterDisk / theminor)² / themajor,
  Style["Disk Gravity (default is 1/2 moon)", {Blue, 12, Bold}]}, 0, 10 000 000},
Style["           (same density ⇒ M_disk = diskthickness/themajor (MaxOuterDisk/theminor)²) M_moon", {Blue, Italic}],

{{diskthickness, 0.02, "Disk thickness"}, 0.02, 1},
```

```
{{theminor, 0.999, "Minor axis of spheroid (must be <1)"}, 0, 0.999},
{{themajor, 1, "Major axis of spheroid"}, theminor, 2},
{{negratio, 2.5, Style["Ratio of excised radius to spheroid radius", {Bold, Darker[Red]}]}, 1, 100},
{{theomega, 15, Style["Rotation Rate, ω", {Bold, Darker[Red]}]}, 0, 60},
Delimiter,
{{plotmin, 0, "Minimum radius plotted"}, 0, 100},
{{plotradius, 6, "Range of radii plotted"}, 0.4, 100},
{{mincontour, 33, "Minimum contour plotted"}, 0, 10 000},
{{numcontours, 100, Style["Separation between contours", Bold]}, 1, 1500},
{{maxcontour, 50 000, "Maximum contour plotted"}, 10 000, 250 000}
]
```

Out[45]=

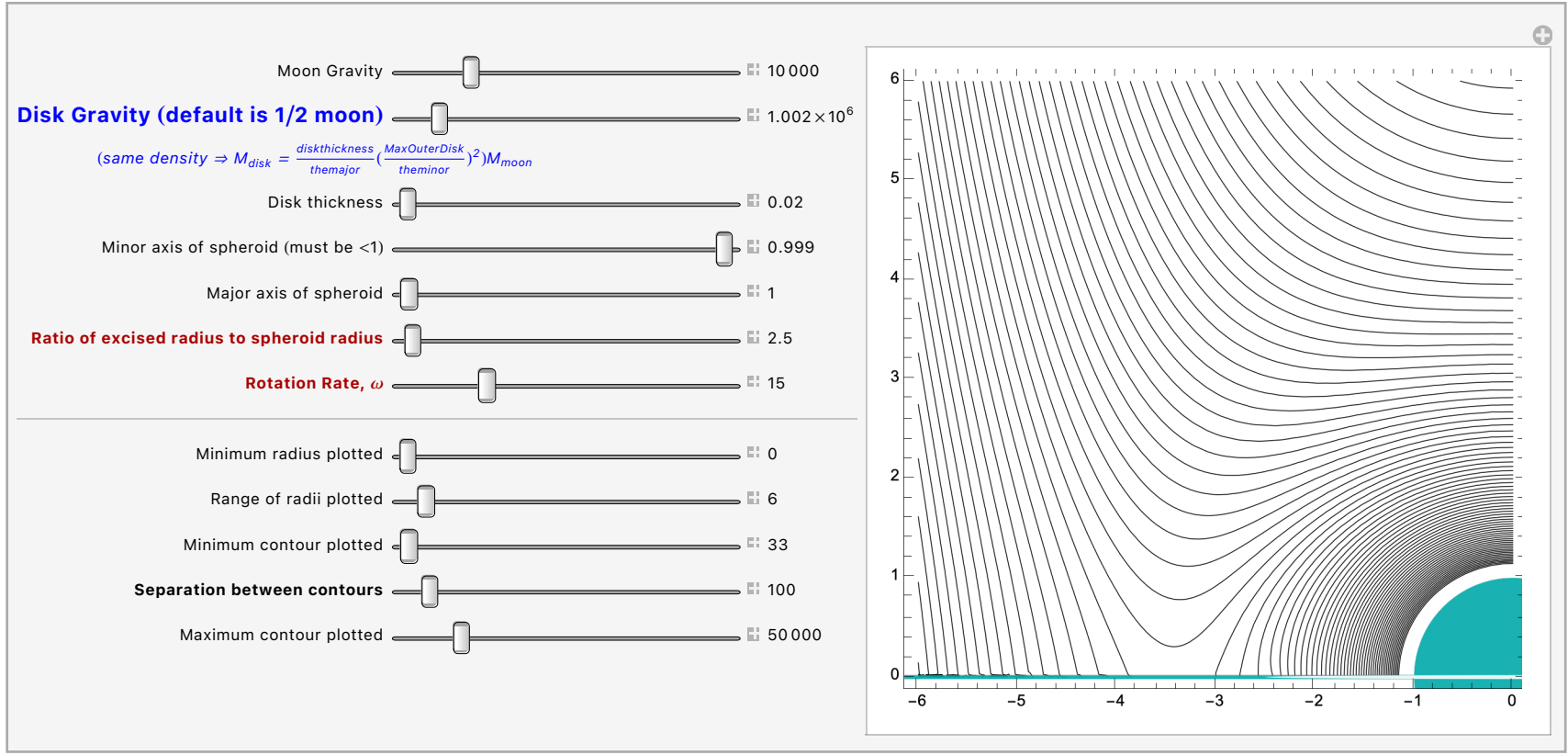